# FPGA Based Sinusoidal Pulse Width Modulated Waveform Generation for Solar (PV) Rural Home Power Inverter

S. N. Singh, A. K. Singh

**Abstract-**With the increasing concern about global environmental protection and energy demand due to rapid growth of population in developing countries and the diminishing trend of resources of conventional grid supply, the need to produce freely available pollution free natural energy such as solar/wind energy has been drawing increasing interest in every corner of the world. In an effort to utilize these energies effectively through Power converter, a great deal of research is being carried out by different researchers / scientist and engineers at different places in the world to meet the increasing demand of load . The study presents methodology to integrate solar (PV) energy (which is freely available in every corner of the world) with grid source and supplement the existing grid power in rural houses during its cut off or restricted supply period. In order to get consistency in supply a DG is also added as a standby source in the proposed integration of network. The software using novel Direct PWM modulation strategy and its soft control features extend the flexibility to control converter (inverter) parameters like voltage, frequency, number of samples of PWM pulses constituting sine-wave without changing any hardware configuration in the circuit. The system simulation of PWM Pulse generation has been done on a XILINX based FPGA Spartan 3E board using VHDL code. The test on simulation of PWM generation program after synthesis and compilation were recorded and verified on a prototype sample.

**Index Terms**-Total harmonic Distortion (THD); Field Programmable Gate Array (FPGA); Pulse Width Modulation (PWM; Photovoltaic (PV); Maximum Power Point (MPP)

——————————   ◆   ——————————

## 1. INTRODUCTION

THE basic need of an electrical energy is increasing with the rapid growth of population in urban, sub-urban and rural sectors. On the other end, the conventional grid supply in a grid connected area has become standstill due to diminishing trend of raw material resources and its further extension is not possible due to various technical, political and economic reasons. To meet the excess energy demand, alternative renewable energy sources like solar/wind etc with energy storage device i.e. Battery are being used to work as a standalone power source or in sharing mode with Grid or DG power source**.** Among these two sources solar energy is preferred as it is easily available in every part of the country in the world where as wind energy is restricted to the coastal area only. A purely solar power converter, if used alone, may become very expensive as far as initial investment is concerned. Further, due to varying solar insolation, the battery barely gets time to fully charge from a single PV source due to varying sun radiation or from the limited available grid source especially in rural sector to its full extent.
Hence the solar power system needs to be integrated with supplementary additional DG back up sources in order to deliver 24 hour power.

The system can also work as a standalone power source in a grid deprived area in remote rural sectors by adding more number of PV modules and battery bank. The optimal utilization of these sources is possible with efficient smart adaptive Power converter and adopting optimal load management.

In the present study, the Pulse width modulated (PWM) adaptive intelligent Power converter (inverter) has been designed and developed where the input DC power stored in the battery bank, obtained through PV and /or Grid sources, has been digitized to produce a sequence of PWM pulses (approximated to a sine wave) at the output of power converter and deliver power to the load. The traditional analog Sine-Triangular method for generating PWM pulses adopt the technology where a high frequency carrier signal is compared with sinusoidal wave as reference signal, set at desired output frequency [1], and thus needed two signals to produce PWM signal (Fig.1.).

————————————————

- *S. N. Singh is with the Department of Electronics Engineering, National Institute of Technology, NIT, Jamshedpur, India.*
- *A. K. Singh is with the Department of Electrical Engineering, National Institute of Technology, NIT, Jamshedpur, India.*

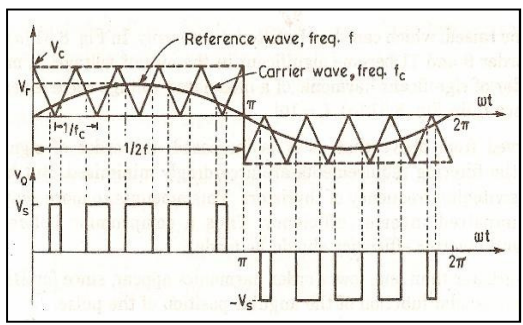

Fig. 1. Generation of PWM Pulses by Sine- Triangular Method

In the present scheme, the PWM pulses are directly generated using a new technique through software program coded with VHDL and downloaded in FPGA Spartan 3E starter kit to produce base drive signals for inverter power device switches. The FPGA VLSI technology offers a fast system with many more advantages as compared to other conventional technology including DSP based controller etc. The software program can easily be changed to optimize and control the inverter parameters like frequency, voltage amplitude, number of PWM pulses in half cycle etc. without changing the hardware circuit. The PWM output waveform designed with high number of PWM pulses in a half cycle can produce a low value of THD content (less than 3-5% THD) and approximate very near to a sine wave which is comparable with the quality of the sine wave of the grid supply.

## 2. SYSTEM CONFIGURATION

The block schematic diagram of FPGA based PWM solar inverter is shown in Fig.2.

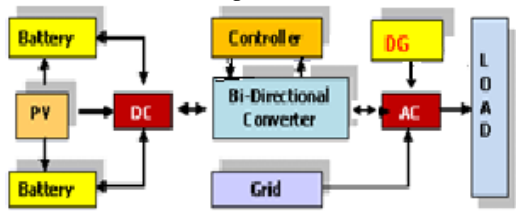

Fig. 2. Block Diagram of Solar PWM Inverter Power supply system

The system works under three modes of operation namely:
- Charging Mode (PV/Grid during sun-hour/available Period )
- Inverting Mode (Grid Cut off or restricted (load shedding)Period )
- Optimally controlled DG operation (during the period when the system does not support to deliver power to load)

In the charging mode, the input energy obtained through PV source during sun- hour is stored in battery to a level more than 12V till it reaches cut-off limit of 13.4V. The charging is also shared proportionately with the grid/DG source using time regulator circuit through sinusoidal PWM bi-directional inverter, if needed during the low radiation/cloudy period, to meet the required end user load energy demand.

The status of input energy resources (i.e. PV/Battery/DG) and load power are depicted in Table 1.

TABLE 1
STATUS OF INPUT AND LOAD POWER

| PV/Battery | Grid | DG | Battery Status | Load power |
|---|---|---|---|---|
| 1 | 1 | 0 | Charging | Grid |
| 0 | 1 | 1 | Charging | Grid |
| 1 | 0 | 0 | Discharging | Battery |
| 0 | 0 | 1 | Charging | DG |

## 3. ADAPTIVE POWER CONTROLLER

The system works on integration of input Power sources (i.e. PV, Battery) with the Grid/DG Power sources and delivers a consistent Power to varying Load(s) as per user demand. The adaptive power control action is governed by Energy balance equation as follows

Load Power = (PV Power / Battery Power  (1)
+ Grid /DG Power)

The status of power flow and logical adaptive control action is shown in Table 2.

TABLE 2
STATUS OF INPUT POWER SOURCES AND LOGICAL OUTPUT

| Energy Resources | Status | Logical output |
|---|---|---|
| PV | 12V or above | 1 |
|  | < 12V | 0 |
| Battery | 13.8V - 10.1V | 1 |
|  | Beyond this limit on higher or lower side | 0 |
| Grid/DG | 220V+ / - 10% | 1 |
|  | Less | 0 |

## 4. SOLAR POWER REGULATOR

The PV sources deliver variable power which depends on solar radiation which varies from 200-1000W/m² during sun hour period. In order to extract/draw maximum power from PV source, solar power regulator has been incorporated which regulates the voltage to respective Maximum Power Point (MPP) i.e. the intersection point between load line and V-I curve at different insolation(radiation) level as shown in fig. 3. This is achieved through intelligent controller unit of solar (PV) converter system.

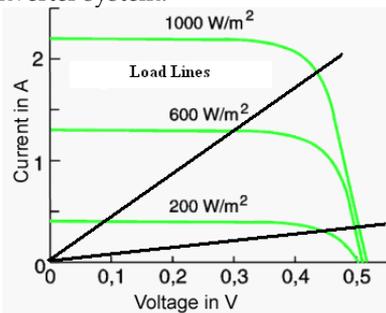

Fig. 3. V - I Characteristic of a Solar (PV) Cell

## 5. DESIGN ALGORITHM: PV MODULE & BATTERY BANK SIZING

The PV module and battery bank are designed [2,3,4] to meet the energy requirement of high priority assigned loads in standalone mode. The optimal sizes are computed on the basis of average daily load requirement. The load requirements are accessed on daily and monthly basis throughout the year and average daily power requirement is computed as per the expression (equation 2) given below

$$P_{daily} = \sum_{i=1}^{i=24} P_i \qquad (2)$$

Where i = 1, 2....24 Hour (Hr)

The PV size is determined by equation 3 and 4 using energy balance optimization technique. The availability of power obtained from PV cell during sun hour period (which is normally taken as 6.2 hour) is matched with daily load power requirement.

$$P_{PV} \times \text{Sun Hour} = P_L \text{ (Watt-hour)} = \sum_{i=1}^{i=24} P_i \qquad (3)$$

i.e. $P_{PV} \text{ (Watt)} = \dfrac{P_L}{\text{Sun hour}} \qquad (4)$

Similarly Battery size is also calculated by equation 5 on the basis of power rating of critical loads and energy requirement during deficit of PV power and grid cut-off period (equation5)

$$\text{Battery AH} = \dfrac{\text{Critical Load Power x Hr}}{\text{Battery Voltage}} \qquad (5)$$

## 6. DESIGN PARAMETER

Design parameters of load, energy sources and system parameters are given in Table 3, Table 4 and Table 5.

TABLE 3
LOAD POWER REQUIREMENT

| Electrical Home Appliances | Load Hour@24Hour |
|---|---|
| Light | 300Wh |
| Fan/TV | 700Wh |
| Pump | 800Wh |

TABLE 4
DESIGN AND SPECIFICATON OF PROTOTYPE INVERTER SYSTEM

| Energy Parameter(s) | |
|---|---|
| Critical Load(s) | 1800Wh Over @24Hr |
| PV | 75Wp |
| Battery | 12V ,150 Ah (2X80Ah) |

TABLE 5
SPECIFICATIONS OF SOLAR POWER CONVERTER SYSTEM

| System Specification | | |
|---|---|---|
| PV array voltage/Power | | 12V / (75W) |
| Battery voltage and capacity | | 12V,150Ah (4x40Ah) |
| Charging/ Discharging current(A) | | 1 – 20 Amp |
| Load(W) | (Resistive) | TV 100W |
| | (Capacitive) | CFL: 18W |
| | (Inductive) | Motor Pump(1/4HP) |
| Inverter (Prototype) | Input Output Power rating Frequency | 12-0-12 / 220±10%, 300VA 50 Hz |
| | Grid (utility) | 220V ± 20%, 50Hz |
| | Power switching Devices | 5*2N3055Trasistor |
| | Generator | 220V, 50 Hz, 750VA |

## 7. INVERTER TOPOLOGY

Inverter is a device which converts the DC power source into AC power a centre-tap inverter topology has been configured to generate PWM sine wave pulses as shown in Fig .4. The semiconductor Power switches may be controlled to produce a multilevel three output voltage state:

$V_o = + V_{dc}$  ( T1 closed )
$V_o = - V_{dc}$  ( T2 closed )
$V_o = 0$      ( Transition from $V_{cc}$ to $- V_{cc}$)

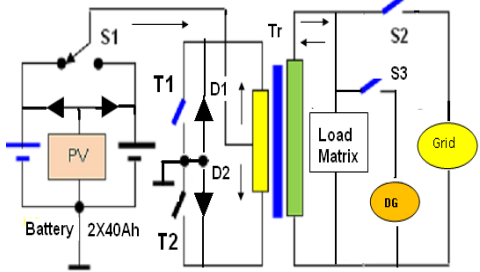

Fig. 4. Circuit Diagram of Power Inverter

The main switching signal (MSS) and polarity control signals (PCS) for N number of PWM base drive pulses is generated by software program. The PWM control base drive gate signals switch on and off the positive group and negative group of inverter power switching devices and thus produces PWM AC pulses approximated to sine wave as shown in Fig. 5.

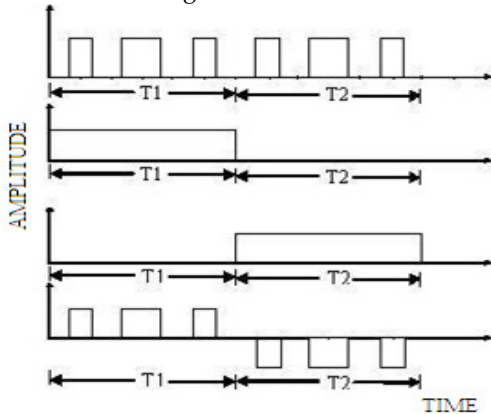

Fig. 5. PWM switching Pulse Generation (Y axis=Amplitude, X axis=Time (T1=T2=10ms i.e. 50Hz)

## 8. METHODOLOGY

The PWM pulses as shown in fig.6 are generated by the software using following mathematical expression (equation 6) [5,6]

$$P_i = K \frac{180}{2N} \times \left[ 2 \sin(2i-1) \frac{\pi}{2N} \right] \qquad (6)$$

Where i = 1 …N (number of PWM pulses)
$P_i$ = Pulse width of PWM pulses
K= (Voltage Regulating Factor (0-1)

The switching angles for the trailing and falling edges are computed by equation 7 and 8.

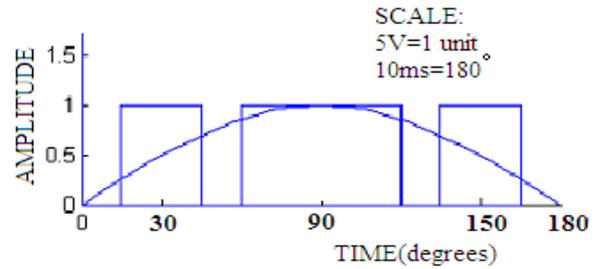

Fig. 6. PWM Wave form Generation

## 9. COMPUTATION OF TIMING OF PWM PULSES

The Pulse width of PWM pulses can be computed using Direct Modulation strategy (equation 7)

$P_i = (180/2N) * 2 \sin (2*i-1) * (\pi /2N)$  (7)
Where,
    i = 1........N (Number of PWM pulses per half cycle of approximated sine wave

The values of switching angles (equation 8 and 9) corresponding to different values of N was calculated from the formula given below:

1. For rising edge of PWM pulses
   $\square = 180 / 2N - 180 * \sin (2*i-1) * \pi /4N$  (8)
2. For falling edge of PWM pulses
   $\square = 180 / 2N + 180 * \sin (2*i-1) * \pi /4N$  (9)

The switching time corresponding to above angles have been computed using MATLAB code to get the values of switching angles. The pre-computed timing of PWM switching pulses (N=3) approximated to a half sine-wave as shown above in Figure 6 for output frequency of 50Hz (equation 10) is shown in Table 6.

Total time (Half cycle sine wave)
   $= (\sum_{i=1}^{N+1} T_i + \sum_{i=1}^{N} P_i)$  (10)
   = 10 ms
i.e. Output frequency = 50Hz

TABLE 6
PWM SWITCHING TIMING PULSES (N=3)

| Notch Width (ms) | Pulse Width (ms) |
|---|---|
| $T_1$ = 0.833 | P1 = 1.667 |
| $T_2$ = 0.833 | P2 = 2.217 |
| $T_3$ = 1.950 | P3 = 1.667 |
| $T_4$ = 0 .833 | |

## 10. EXPERIMENTAL INVESTIGATION

The software used to design the system is XILINX 9.2i. It is a vast software mainly used in industries for designing, testing and development of digital ASIC's. The software process of the system is basically divided into three stages:
- Design
- Implementation
- Simulation

During the design stage shown, the VHDL code for the generation of SPWM waveform from the FPGA has been written *(Appendix-I)*.

The software program of PWM generation was downloaded in FPGA Spartan 3E board using Verilog hardware descriptive language (VHDL) codes. The result of simulated wave using inbuilt ISE simulator is shown in Fig7.

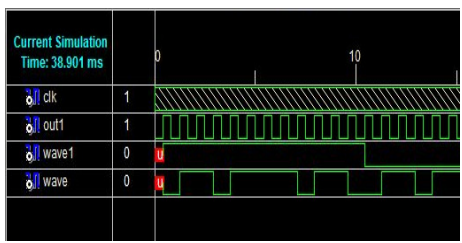

Fig. 7. PWM Simulated Waveform

The hardware implementation has been done through FPGA Spartan 3E Board Fig7. The integration of FPGA controller with Power converter produces PWM waveform at its output. The waveform of PWM control pulses and the approximated sine-wave produced by inverter across load is recorded in oscilloscope as shown in Fig 9 and 10.

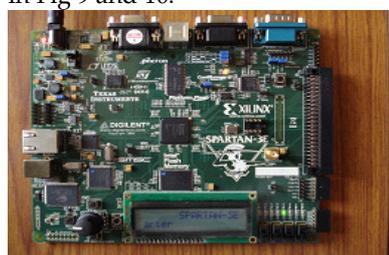

Fig. 8. FPGA Spartan 3 E Board

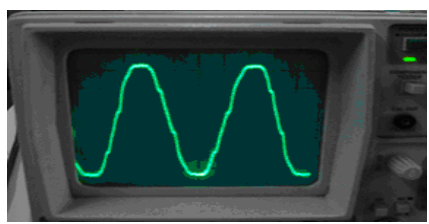

Fig. 9. Oscilloscopic image of MSS (Top) and PCS (Bottom) of PWM Wave form for N=3 (i.e. Number of Pulses in a half cycle )

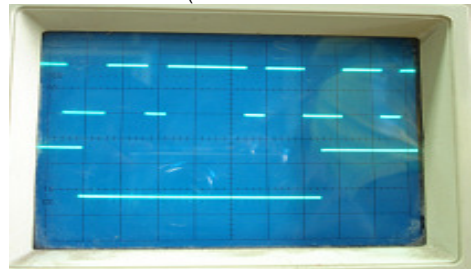

Fig. 10. Load Waveform

## 11. LOAD MATRIX

The load matrix network is switched on sequentially through intelligent controller such that load power can be matched with available power to be drawn from integrated PV, Battery and Grid/DG input sources. Thus power converter feed power to various household and agro based loads in rural areas as per their demand and priority fixed by the users.

The priority of switching of loads is set by users to manage the peak load power requirement over the period of its operation and sequentially switched on one by one from high priority to low priority loads till total summed up load power matches the output power available from input power sources (i.e. PV, Battery and/or Grid/DG power). The reverse sequential process takes place in case power reduces in all or any of the input sources.

## 12. HARMONIC ANALYSIS

The harmonic analysis was carried out using MATLAB (Version-7) software. The computed value of THD express in % for N number of PWM pulses is shown in Figure 11.

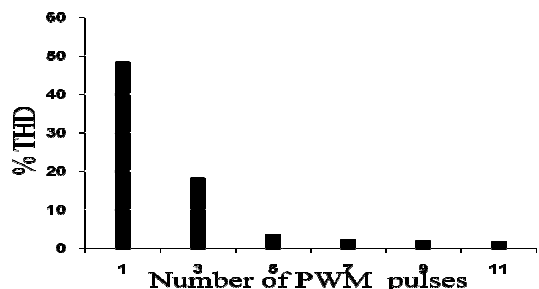

Fig. 11. Harmonic Analysis (X axis =Number of PWM pulses in a half cycle of approximated sine wave(N) , Y axis = % THD )

## 13. EFFICIENCY

The efficiency of Inverter has been observed as almost constant as shown in Fig 12.

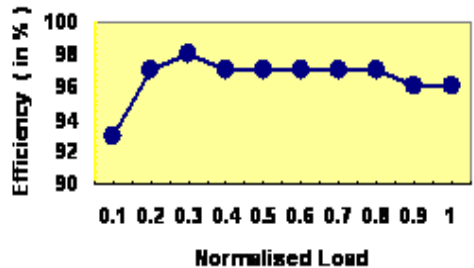

Fig. 12. Plot of Efficiency (in %) Vs Load (in %) of PV converter System

## 14. CONCLUSION

A hybrid PV converter system back up with standby DG source proposed in this study for rural home and agro-based load applications is considered as appropriate design for developing countries like India where grid cut-off happens very frequently and does not provide the house with any back up or sufficient safe back up supply. The performance of the proposed system is better and its software controlled features offer more flexibility as compared to other PV converter system. Although DG has been integrated in the scheme, but it is used only under the rare conditions like cloudy days, low sun radiation as well as long duration of grid failure. Thus, an uninterrupted power supply of 24 hours can be obtained using the proposed scheme.

The use of FPGA technology to generate PWM pulses for solar inverter using VHDL programming language has successfully been implemented in the present study. The software controlled program can alter the inverter parameter(s) and can easily be outputted through FPGA board. The grid interactive PV inverter can also be used as a stand-alone Power supply by adding more number of PV module as per load requirement in a grid deprived area especially in rural sector.

The solar inverter finds its application in the following areas.

- Power supply (UPS)
- Supplementary source of Grid supply
- Rural Industry
- Solar power houses
- Agro-based equipment
- Domestic electrical appliance
- Water pumping system
- Telecommunication system etc.

**Dr S.N.Singh** had completed doctoral Ph.D degree at the Department of Electrical Engineering, National Institute of Technology Jamshedpur (India) in 2009. He completed his Master's degree in Electrical Engineering from Ranchi University (India) with specialization in Power Electronics in1991. He obtained B.Tech degree in Electronics and communication engineering from BIT Mesra, Ranchi - Jharkhand (India) (A Deemed University) in 1979. Presently his area of interest is in solar energy conversion technology. He had published more than 10 papers in National and International journals based on his research work. He had carried out consultancy and R&D work in industry and developed software and carried out innovative project work there. He had remained *Head of Department of Electronics Engineering* for two terms and presently heading VLSI Project as a *Co-coordinator* of VLSI Project (SMDP-II) sponsored by Ministry of Information Technology, Government of India and Prof-in charge of several administrative and academic committees including *Chief Warden* of the Institute. He had total *30 years of experience* including administrative, research and industrial experience at executive level in different industries and CSIR (Govt. of India) lab and presently *a senior faculty* in the Department of Electronics Engineering in National Institute of Technology (An Autonomous Institution of MHRD, Govt. Of India) Jamshedpur (India).

**Dr A.K.Singh** received his Ph.D degree in Electrical Engineering from Indian Institute of Technology, Kharagpur(India) in 1995. He obtained his M.Tech degree from BHU University (India) and B.Tech degree from National institute of Technology Kurushetra (India). Presently he is the *Head of Department of Electrical Engineering* at National Institute of Technology, Jamshedpur (An educational Institute of MHRD, Govt of India) Jamshedpur (India). His area of interest is in Electrical Control System and Utilization of Renewable Solar Energy Sources. He had published more than 10 papers in National and International journal in the research area. He had remained Prof-in Charge in several administrative and academic committee of Institute. Presently under him two Ph.D research fellows are pursuing their research work in the field of intelligent control system. He had more than *30 years of experience* in academic, administrative, research and consultancy work. He had guided several professional engineers during their M.Tech study and developed software and carried out innovative project work in industry.

**Appendix-I**

**Sinusoidal PWM Pulse Generation For N=3**

**VHDL Code :**
---------------------------------------------------------------------------------------------------------------------------------------------------

```
library IEEE;
use IEEE.STD_LOGIC_1164.ALL;
use IEEE.STD_LOGIC_ARITH.ALL;
use IEEE.STD_LOGIC_UNSIGNED.ALL;
entity N3 is
    Port ( CLK : in  STD_LOGIC;
           SEL : buffer  STD_LOGIC;
           PWM : out  STD_LOGIC;
           WAVE1 :buffer  STD_LOGIC);
end N3;

architecture Behavioral of N3 is
type STATE is (S1,S2,S3,S4,S5,S6,S7);
SIGNAL PR,NX:STATE;
SIGNAL TM:INTEGER RANGE 0 TO 16667;
begin
PROCESS(CLK)
VARIABLE COUNT1:INTEGER RANGE 0 TO 5;
BEGIN
IF(CLK'EVENT AND CLK='1')THEN
COUNT1:=COUNT1+1;
IF(COUNT1=5)THEN
WAVE1<= NOT WAVE1;
COUNT1:=0;
END IF;
END IF;
END PROCESS;

PROCESS(CLK)
VARIABLE count2:INTEGER RANGE 0 TO 500000;
begin
IF(CLK'EVENT and CLK='1') THEN
        count2:=count2+1;
            IF(count2=500000) THEN
            SEL<=NOT SEL;--20ms
            count2:=0;
        END IF;
END IF;
END PROCESS;

PROCESS(WAVE1)
VARIABLE COUNT :INTEGER RANGE 0 TO 16667;
BEGIN
IF(WAVE1'EVENT AND WAVE1='1')THEN
COUNT:=COUNT+1;
IF(COUNT=TM)THEN
PR<=NX;
COUNT:=0;
```

```
END IF;
END IF;
END PROCESS;
PROCESS(PR)
BEGIN
CASE PR IS
WHEN S1=>
                PWM<='0';
                TM<=4167;
                NX<=S2;
WHEN S2=>
                PWM<='1';
                TM<=8333;
                NX<=S3;
WHEN S3=>
                PWM<='0';
                TM<=4167;
                NX<=S4;
WHEN S4=>
                PWM<='1';
                TM<=16667;
                NX<=S5;
WHEN S5=>
                PWM<='0';
                TM<=4167;
                NX<=S6;
WHEN S6=>
                PWM<='1';
                TM<=8334;
                NX<=S7;
WHEN S7=>
                PWM<='0';
                TM<=4167;
                NX<=S1;
END CASE;
END PROCESS;

end Behavioral;
```